\documentclass[letterpaper,twocolumn,10pt]{article}

\usepackage{usenix,epsfig,endnotes}
\usepackage{shadow}        
\usepackage{hhline}        
\usepackage{tabularx}      
\usepackage{latexsym}      
\usepackage{pslatex}       
\usepackage{booktabs}      
\usepackage{multirow}
\usepackage{url}           
\usepackage{breakurl}

\usepackage{balance}       
\usepackage{cite}          
\usepackage{amsmath}
\usepackage{rotating}
\usepackage{xspace}
\usepackage{color}
\usepackage[usenames,dvipsnames,svgames,table]{xcolor}
\usepackage{colortbl}
\usepackage{multicol}
\usepackage{nicefrac}
\usepackage{enumitem}
\setlist{nolistsep}
\usepackage{fixltx2e}
\usepackage{float}

\usepackage{algorithm}
\usepackage[noend]{algpseudocode}
\usepackage{xpatch}

\makeatletter
\def\BState{\State\hskip-\ALG@thistlm}
\makeatother

\algdef{SE}[VARIABLES]{Variables}{EndVariables}
   {\algorithmicvariables}
   {\algorithmicend\ \algorithmicvariables}
\algnewcommand{\algorithmicvariables}{\textbf{global variables}}

\algnewcommand\algorithmicforeach{\textbf{for each}}
\algdef{S}[FOR]{ForEach}[1]{\algorithmicforeach\ #1\ \algorithmicdo}

\algnewcommand{\LeftComment}[1]{\Statex \(\triangleright\) #1}

\usepackage[T1]{fontenc}
\usepackage{textcomp}

\newcommand{\changed}{\textcolor{black}}

\newcommand{\bolt}{Bolted}
\newcommand{\hil}{HIL}
\newcommand{\bmi}{BMI}
\newcommand{\keyl}{Keylime}
\newcommand{\linuxboot}{LinuxBoot}
\newcommand{\llinuxboot}{LinuxBoot}
\newcommand{\lhil}{Hardware Isolation Layer}
\newcommand{\lbmi}{Bare Metal Imaging}
\newcommand{\lkeyl}{Keylime}

\usepackage[compact]{titlesec}
\titleformat{\section}
{\sffamily\Large\bfseries}{\thesection}{1em}{}
\titleformat{\subsection}
{\sffamily\large\bfseries}{\thesubsection}{1em}{}
\titleformat{\subsubsection}
            {\sffamily\normalsize\bfseries}{\thesubsubsection}{1em}{}

\ifx\pdfoutput\undefined
\usepackage{graphicx}
\else

\usepackage{graphicx}
\usepackage{epstopdf}
\fi

\usepackage[%
  font=small,
   hypcap=true,
  labelfont=bf
  ]{caption}
\usepackage[%
  hypcap=true
]{subcaption}
\usepackage[%
  pdftex,              
  colorlinks=true,     
  urlcolor=blue,       
  filecolor=blue,      
  linkcolor=black,     
  citecolor=black,     
  pdftitle={},
  pdfauthor={},
  pdfsubject={},
  pdfkeywords={},
  pdfproducer={pdfLaTeX},
  pdfpagemode=UseOutlines,  
  bookmarksopen=true   
]{hyperref}

\usepackage{enumitem}
\setlist{nolistsep}
\newcommand{\fixfig}{\vspace{-0.8ex}}

\begin{document}

\graphicspath{ {figs/} }

\title{\Large \bf Supporting Security Sensitive Tenants in a Bare-Metal Cloud\thanks{Mosayyebzadeh, Mohan, and Tikale contributed equally.}
\thanks{{\scriptsize DISTRIBUTION STATEMENT A. Approved for public release. Distribution is unlimited. This material is based upon work supported by the Under Secretary of Defense for Research and Engineering under Air Force Contract No. FA8702-15-D-0001. Any opinions, findings, conclusions or recommendations expressed in this material are those of the author(s) and do not necessarily reflect the views of the Under Secretary of Defense for Research and Engineering.}}}

\author{\rm Amin Mosayyebzadeh$^{1}$, \rm Apoorve Mohan$^{4}$, \rm Sahil Tikale$^{1}$,\\
    \rm Mania Abdi$^{4}$, \rm  Nabil Schear$^{2}$, \rm Charles Munson$^{2}$, \rm Trammell Hudson$^{3}$\\
    \rm Larry Rudolph$^{3}$, \rm Gene Cooperman$^{4}$, \rm Peter Desnoyers$^{4}$, \rm Orran Krieger$^{1}$\\
\small {\em  $^1$Boston University \quad
  $^2$MIT Lincoln Laboratory  \quad
   $^{3}$Two Sigma \quad
  $^{4}$Northeastern University}
}

\date{}

\maketitle

\begin{abstract}


\bolt\ is a new architecture for bare-metal clouds that enables tenants to control tradeoffs between security, price, and performance. Security-sensitive tenants can minimize their trust in the public cloud provider and achieve similar levels of security and control that they can obtain in their own private data centers.  At the same time, Bolted neither imposes overhead on tenants that are security insensitive nor compromises the flexibility or operational efficiency of the provider. Our prototype exploits a novel provisioning system and specialized firmware to enable elasticity similar to virtualized clouds.  Experimentally we quantify the cost of different levels of security for a variety of workloads and demonstrate the value of giving control to the tenant. 
\end{abstract}

\section{Introduction}

There are a number of security concerns with today's clouds. First, virtualized clouds collocate multiple tenants on a single physical node, enabling malicious tenants to launch side-channel and covert channel attacks~\cite{ristenpart2009hey,Kocher2018spectre,Lipp2018meltdown,liu,razavi2016flip, attackcache_2017, coreside_2017} or
exploit vulnerabilities in the hypervisor to launch attacks both on tenants running on the same node~\cite{king2006subvirt, Perez-Botero:2013} and on the cloud provider itself\cite{sze_hardening:2016}. Second, popular cloud management software like OpenStack can have a  trusted computing base (TCB) with millions of lines of code and a massive attack surface\cite{DBLP:journals/iacr/HoganMRCDHVZ18}.  Third, for operational efficiency, cloud providers tend to support {\em one-size-fits-all} solutions, where they apply uniform solutions (e.g. network encryption) to all customers; meeting the specialized requirements of highly security sensitive customers may impose unacceptable costs for others. Finally, and perhaps most concerning, existing clouds provide tenants with very limited visibility and control over internal operations and implementations; the tenant needs to fully trust the non-maliciousness and competence of the provider.

While bare-metal clouds~\cite{Softlayer:2015,RackSpace:2015,Internap:2015,Packet.net,AWSBare} eliminate the security concerns implicit in virtualization, they do not address the rest of the challenges described above. For example, OpenStack's bare-metal service still has all of OpenStack in the TCB.  As another example, existing bare-metal clouds ensure that previous tenants have not compromised firmware by adopting a one-size-fits-all approach of validation/attestation or re-flashing firmware.  The tenant has no way to programmatically verify the firmware installed and needs to fully trust the provider.  As yet another example, existing bare-metal clouds require the tenant to trust the provider to scrub any persistent state on the physical machine before allocating the machine to other tenants.\footnote{See, for example, IBM Cloud's security policy for scrubbing local drives here https://tinyurl.com/y75sakn4. Note that scrubbing local disks can require hours of overhead on transferring computers between tenants; dramatically impacting the elasticity of the cloud.}

These issues are a major concern for ``security-sensitive'' organizations, which we define as entities that are both willing to pay a significant price (dollars and/or performance) for security and that have the expertise, desire, or requirement to trust their own security arrangements over those of a cloud provider.  Many medical, financial and federal institutions fit into this category.  Recently, IARPA, who represents a number of such entities, released an RFI~\cite{IARPA_rfi_2017} that describes their requirement for using future public clouds; to
``\emph{replicate as closely as possible the properties of an air-gapped private enclave}''  of physical machines. More concretely\footnote{Per private communications with RFI authors.} this means a cloud where the tenant trusts the provider to make systems available but where confidentiality and integrity for a tenant's {\em enclave} is under the control of the tenant who is free to implement their own specialized security processes and procedures.

By our definition the majority of computing demands are not highly security-sensitive, thus providing a high-security option within a commercially-viable future cloud must not impact the efficiency of providing service to other tenants. Is this possible?  Can we make a cloud that is appropriate for even the most security sensitive tenants? Can we make a cloud where the tenant does not need to fully trust the provider? Can we do this without performance impact on tenants who are happy with the security levels of today's clouds?

The \bolt\ architecture and prototype implementation, described in this paper, demonstrates that the answer to these questions is ``yes.''  The fundamental insight is that to implement a bare metal cloud only a minimum {\em isolation service} need to be controlled by the provider; all other functionality can be implemented by security-sensitive tenants on their own behalf, with provider-maintained implementations available to tenants with more typical security needs.

\bolt\ defines a set of micro-services, namely an {\em isolation service} that uses network isolation technologies to isolate tenants' bare-metal servers, a {\em provisioning service} that installs software on servers using network mounted storage, and an {\em attestation service} that compares measurements (hashes) of firmware/software on a server against a whitelist of allowed software.  All services can be deployed by the provider as a one-size-fits-all solution for the tenants that are willing to trust the provider.

Security sensitive tenants can deploy their own provisioning and attestation service thereby minimizing their trust in the provider.
The tenant's own software executing on machines (already trusted by the tenant), can validate measurements of code to be executed on some newly allocated server against her expectations rather than having to trust the provider. Further, the tenant's attestation service can securely distribute keys to the server for network and disk encryption.  Using the default implementation of \bolt\ services, a tenant's enclave is protected from previous users of the same servers (using hardware-based attestation), from concurrent tenants of the cloud (using network isolation and encryption), and from future users of the same servers (using network mounted storage, storage encryption, and memory scrubbing).   Further, a tenant with specialized needs can modify these services to match their requirements; the provider does not sacrifice operational efficiency or flexibility for security-sensitive customers with specialized needs since it is the tenant and not the provider responsible for implementing complex policies.

{\bf Key contributions of this paper are:}

An {\bf architecture} for a bare-metal cloud that: 1) enables security-sensitive tenants to control their own security while only trusting the provider for physical security and availability while 2) not imposing overhead on tenants that are security insensitive and not compromising the flexibility or operational efficiency of the provider.  Key elements of the architecture are: 1) disk-less provisioning that eliminates the need to trust the provider for disk scrubbing (as well as the huge cost), 2) remote attestation (versus validation or re-flashing) to provide the tenant with a proof of the firmware and software running on their server and 3) secure deterministically built firmware that allows the tenant to inspect the source code used to generate the firmware.


\changed{A {\bf prototype implementation} of the \bolt\ architecture where all its components are made available by us open-source, including the isolation service (\lhil~\cite{hil,github_hil_2018}), a deterministic Linux-based minimal firmware (\linuxboot~\cite{linuxboot_2018,hudson_linuxboot_2018}, a disk-less bare-metal provisioning service (\lbmi~\cite{bmi, github_malleable_2018}), a remote attestation service (\lkeyl~\cite{keylime, github_python_keylime_2018}), and scripts that interact with the various services to elastically create secure private enclaves. As we will discuss later, only the microservice providing isolation (i.e., \lhil\ ) is in the TCB and we show that this can, in fact, be quite small; just over 3K LOC in our implementation.}

A {\bf performance evaluation} of the \bolt\ prototype that demonstrates: 1) elasticity similar to today's virtualized cloud ($\sim$3 minutes to allocate and provision a physical server), 2) the cost of attestation has a modest impact $\sim$25\% on the provisioning time,
3) there is value for customers that trust the provider in avoiding extra security (e.g.,$\sim$200\% for some applications), while 4) security-sensitive customers can still run many non-IO intensive applications with negligible overhead and even I/O intensive BigData applications with a relatively modest (e.g., $\sim$30\%) degradation.



\section{Threat Model}\label{threat}

We describe the threats to the victim, a tenant renting bare-metal servers from the cloud, and describe approaches taken by \bolt~ to safeguards against them. We consider external entities (hackers), malicious insiders in the cloud provider's organization and all other tenants of the server---both past and future---as potential adversaries to the victim. We assume that the goal of the adversary is to steal data, corrupt data, or deny services to the victim by gaining access to the victim's occupied servers or network.
Our goal is to empower the tenant with the ability to take control of its own security; it is up to the tenant to make the tradeoff decision between the degree to which it relies on the provider's security systems versus the harm that it may suffer from a successful attack.  

The cloud provider is always trusted 
with the physical security of the datacenter, thus any attacks involving physical access to the infrastructure, including power and noise analysis, bus snooping, or decapping chips~\cite{Halderman_coldboot:2008, Szefer_phyattacks:2014, guri_powerhammer:2018} are out of scope of a tenant's control. The provider is also trusted for the availability of the network, node allocation services, and any network performance guarantees. We assume that the cloud itself is vulnerable to exploitation by external entities (hackers) or a malicious insider (e.g., a rogue systems administrator) but we trust the cloud provider's organization to have necessary systems and procedures in place to detect and limit the impact of such events. For example, the provider can enforce sufficient technical separation of duties (e.g., two-person rule) such that a single malicious insider or hacker cannot both re-flash all the node firmware in a data center and change what hashes the provider publishes for attestation, have both physical and logical access to a node, or make unreviewed changes to the provider's deployed software, etc.  Further, we assume that all servers in the cloud are equipped with a Trusted Platform Module (TPM) - a dedicated cryptographic coprocessor required for hardware-based authentication ~\cite{TPM_trusted_2008}.

We categorize the threats that the tenant faces into the following phases:



\textbf{Prior to occupancy:} Malicious (or buggy) firmware can threaten the integrity of a server, as well as that of other servers it is able to contact.  A tenant server's firmware may be infected prior to the tenant using it, either by the previous tenant (e.g., by exploiting firmware bugs) or by the cloud provider insider (e.g., by unauthorized firmware modification).  If a server is not sufficiently isolated from potential attackers there is also a threat of infection between the time it is booted until it is fully provisioned and all defenses are in place.




\textbf{During occupancy:} Although many side-channel attacks are avoided by disallowing concurrent tenants on the same server, if the server's network traffic is not sufficiently isolated, the provider or other concurrent tenants of the cloud may be able to launch attacks against it or eavesdrop on its communication with other servers in the enclave. Moreover, if network attached storage is used (as in our implementation) all communication that is not sufficiently secured between server and storage may be vulnerable. Finally, there is a threat to the tenant from denial of service attacks.




\textbf{After occupancy:} Once the tenant releases a server,  the confidentiality of a tenant may be compromised by any of its state (e.g, storage or memory) being visible to subsequent software running on the server.




\section{Design Philosophy}
\label{phil}


The key goals of \bolt\ are:
(1) to minimize the trust that a tenant needs to place in the provider,
(2) to enable tenants with specialized security expertise to implement the functionality themselves, and
(3) to enable tenants to make their own cost/performance/security tradeoffs
-- in bare-metal clouds. These goals have a number of implications in the design of \bolt.



First, \bolt\ differs from existing bare metal offerings in that most of the component services that make up \bolt\ can be operated by a tenant rather than by the provider. A security sensitive tenant can customize or replace these services. All the logic that orchestrates how different services are used to securely deploy a tenant's software is implemented using scripts that can be replaced or modified by the user.  Most importantly, the service that checks the integrity of a rented server can be deployed (and potentially re-implemented) by the tenant.

Second, while we expect a provider to secure and isolate the network and storage of tenants, we only rely on the provider for availability and not for the confidentiality or integrity of the tenant's computation. For tenants that do not trust the provider, we assume that \bolt\ tenants will further encrypt all communication between the their servers and between those servers and storage. \bolt\ provides a (user-operated) service to securely distribute keys for this purpose.

Third, we rely on attestation (measuring all firmware and software and ensuring that it matches known good values) that can be implemented by the tenant rather than just validation (ensuring that software/firmware is signed by a trusted party). This is critical for firmware which may contain bugs~\cite{thunderstrike,thunderstrike2,lighteater,atr-smm,x86-harmful,heasman} that can disrupt tenant security. Attestation provides a time-of-use proof that the provider has kept the firmware up to date.
\changed {More generally, the whole process of incorporating a server into an enclave can be attested to the tenant. In addition, the tenant can continuously attest when the server is operating, ensuring that any code loaded in any layer of software (OS, applications and etc., and irrespective of who signed them) can be dynamically checked against a tenant-controlled whitelist.}

Fourth, we have a strong focus on keeping our software as small as possible and making it all available via open source. In some cases, we have written our own highly specialized functionality rather than relying on larger function rich general purpose code in order to achieve this goal. For functionality deployed by the provider, this is critical to enable it to be inspected by tenants to ensure that any requirements are met. For example, previous attacks have shown that firmware security features are difficult to implement bug-free -- including firmware measurements being insufficient~\cite{chronomancy}, hardware protections against malicious devices not being in place~\cite{iommu-attack}, and dynamic root of trust (DRTM) implementation flaws~\cite{wojtczuk2009attacking}.
Further, our firmware is deterministically built, so that the tenant can not only inspect it for correct implementation but then easily check that this is the firmware that is actually executing on the machine assigned to the tenant. For tenant-deployed functionality, small open source implementations are valuable to enable user-specific customization.


Finally, servers are assumed to be stateless with all volumes accessed on-demand over the network.  This removes confidentiality or denial of service attacks by the provider or subsequent tenants of server inspecting or deleting a tenants disk state.  Bare-metal clouds that support stateful servers need to either give the tenant the guarantee that a node will never be preempted (problematic in a pay-for-use cloud model) or ensure that the provider scrubs the disks (trusting the provider and potentially requiring hours with modern disks).  As we will see, stateless servers also dramatically improve the elasticity of the service.


\section{Architecture}
\label{arch}

\bolt\ enables tenants to build a secure enclave of bare-metal servers where the integrity of each server is verified by the tenant before it is allowed to participate in the tenant's enclave.
During the allocation process, a server transitions through the following states: \textbf{free}, or not allocated, \textbf{airlock}, where the integrity of the server is checked, after which it is either \textbf{allocated} to a tenant's secure enclave if it passes the integrity check or \textbf{rejected} if it fails.
In this section, we discuss the \bolt\ components; their operations; the process of server allocation, attestation, and the degrees of freedom in deploying \bolt\ components to support different security requirements and use cases.

\subsection{Components}

\bolt\ consists of four components which operate independently and (in the highest-security and lowest-trust configurations) are orchestrated by the tenant rather than the  provider.

\textbf{Isolation Service:} The Isolation Service exposes interfaces to (de)allocate servers and networks to tenants, and isolate and/or group the servers by manipulating a provider's networking infrastructure (switches and/or routers). Using the exposed interfaces, the servers are moved to \emph{free} or \emph{rejected} state as well -- ensuring the servers are not part of any tenant-owned network. These interfaces are also used to move the servers to the \emph{airlock} state (to verify if they have been compromised) or the \emph{allocated} state (where they are available for the tenant).

The Isolation Service uses network isolation techniques instead of encryption-based logical isolation in order to enforce guarantees of performance and to provide basic protection against traffic analysis attacks. Since the operations performed by these interfaces (on the networking infrastructure) are privileged, the isolation service needs to be deployed by the provider; if a tenant does not trust the provider, it can further encrypt network traffic between their servers.



\textbf{Secure Firmware:} Secure firmware is crucial
towards improving tenant's trust of the public cloud servers; it should consist of following properties.
First, it should be open-source, so that it benefits from large community support in improving its features and fixing any bugs and vulnerabilities. Second, it should be deterministically built so that a tenant can build the firmware from verified source code and independently validate the provider-installed firmware. Third, it must scrub server memory prior to launching a tenant OS -- if the server was preempted from a previous tenant, it must guarantee that the previous tenants` code and data is not present in the memory. Finally, it must provide an execution environment for the attestation agent in the airlock state.

We note that it is challenging to replace computer firmware; even major providers are often forced to install huge binary blobs signed by the hardware manufacturer with no access to the source code. When firmware cannot be replaced, we use the installed firmware for the minimum amount of time in order to download our own secure firmware -- and the servers` pre-installed firmware must support trusted boot~\cite{arbaugh2007trusted}.

While the overall \bolt\ architecture design supports the attestation and security of both system firmware (e.g., BIOS or UEFI) and peripheral firmware (e.g., GPU, network card, etc.), there are no standardized and implemented methods to attest those peripheral firmware to an external party.
Early attempts at standardization are underway, and we expect \bolt\ can leverage them when they mature~\cite{nist800_193}.


\textbf{Provisioning Service:} This service is broadly responsible for three things -- (1) initial provisioning of the server with the software stack (i.e. secure firmware and attestation agent) responsible for its attestation during the \emph{airlock} state, (2) provisioning of the server during the {\em allocated} state (i.e. the server was successfully verified that it was not compromised) with the intended software stack i.e. the operating system and the relevant software packages, and (3) saving and/or deleting the servers` persistent state when a server is released.

The Provisioning Service can be deployed either by the provider or by tenants themselves. The latter option is valuable for security-sensitive tenants who do not want to trust the provider with their operating system images or who want to use their own (e.g., legacy) provisioning systems. The provisioning service must provision the servers in a stateless manner so that the tenants do not have to rely on (and trust) the provider to remove any persistent state after the server is released.

\textbf{Attestation Service:} The Attestation Service consists of two parts: an attestation agent that executes on the server to be attested, and an attestation server that maintains a pre-populated database of known reliable hash measurements of allowed firmware/software (i.e., a whitelist). This service is used during the \emph{airlock} and {\em allocated} states. The Attestation Service can be deployed either by the provider or by the tenant.

During the \emph{airlock} state, the attestation agent (downloaded from the Provisioning Service during initial provisioning) is responsible for sending {\em quotes}\footnote{Hash measurements obtained from and signed by a secure cryptoprocessor such as TPM.} of the firmware and any other software involved during the boot sequence to the attestation server to be matched against the whitelist. Depending on the attestation result obtained from the attestation server, the state of the attested server is changed to allocated or rejected. In the case when the computer firmware cannot be replaced, the trusted boot sequence measurement (until the secure firmware is executed) must be supplied by the provider. Obtaining this measurement is a one-time operation for each server, and this whitelist can be published publicly by the provider.

In the {\em allocated} state, the attestation agent (installed on the tenants' OS) can continuously verify the software stack running against the whitelist present on the attestation server (also referred as {\em Continuous Attestation}). For continuous attestation to work, the software stack should be configured such that it saves new measurements to the cryptoprocessor upon observing any change/modification/access. The attestation agent periodically sends the new hash measurements of software and configuration registered in the cryptoprocessor to the attestation server; if attestation fails (i.e., when any malicious activity is observed), the attestation server alerts the attestation agent.  Continuous attestation protects tenants both against unauthorized execution of executables and against malicious reboots into unauthorized firmware, bootloader, or operating system.
Note that continuous attestation is fundamentally more challenging in a provider-deployed attestation service, as the runtime whitelist (e.g., hashes of approved binaries allowed to be run on the node) must be tenant-generated; we assume continuous attestation is only used by security-sensitive tenants that deploy their own attestation service.



\subsection{Life Cycle}

\begin{figure}[t]
  \centering
	\includegraphics[width=8.5cm]{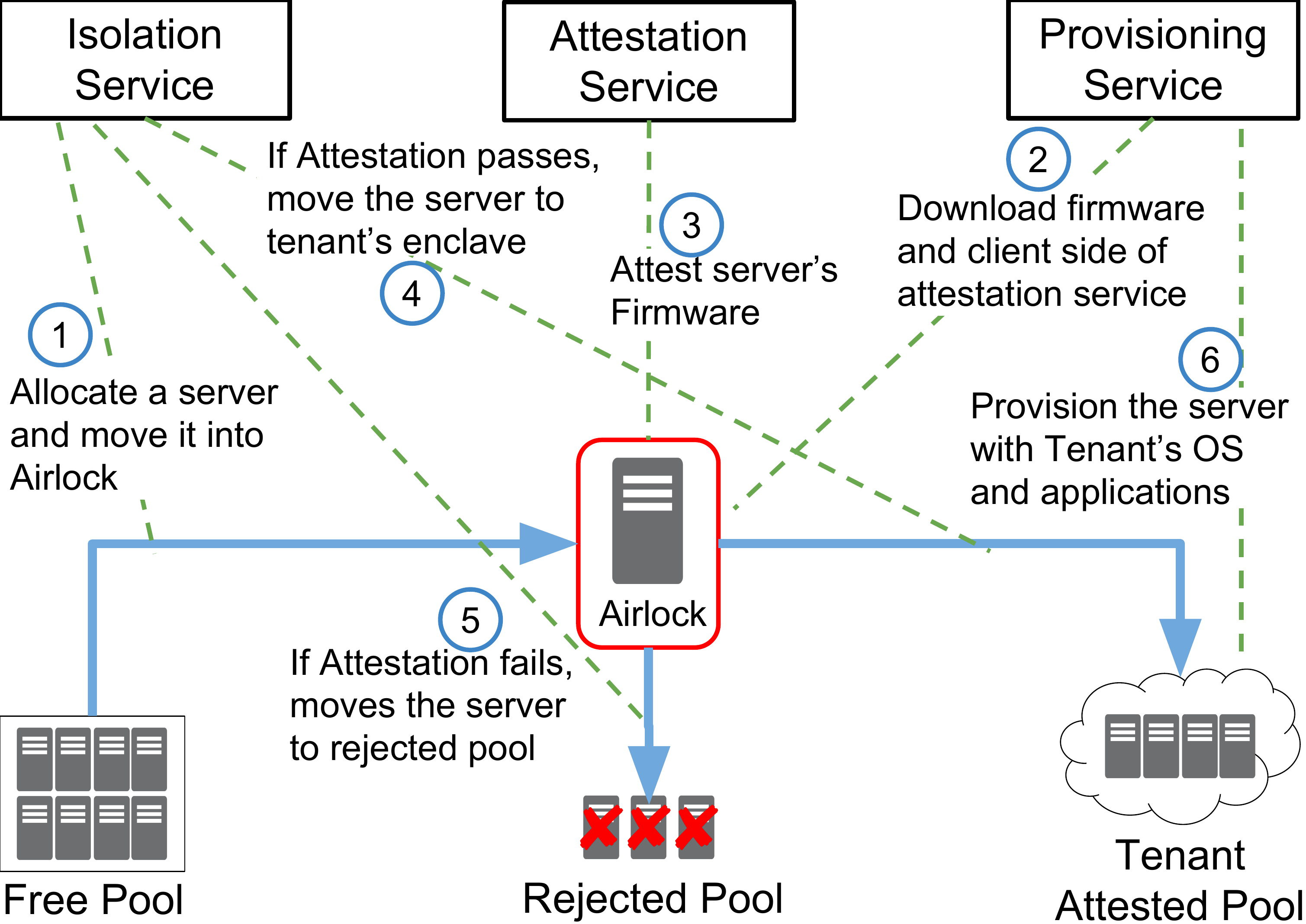}
\fixfig
	\caption {\bolt's Architecture: Blue arrows show state changes and green dotted lines shows the actions during a state change.}
\fixfig
\vspace{-4mm}
	\label{fig:simple}
\end{figure}

The different \bolt\ components do not directly interact with each other, but instead, are orchestrated by user-controlled scripts.
Figure \ref{fig:simple} shows the \textbf{life-cycle of a typical secure server} (in the case of security-sensitive tenant), which progresses through six steps: (\textbf{1}) The tenant uses the Isolation Service to allocate a new bare metal server, create an airlock network, and move the server to that airlock, shared with the Attestation and the Provisioning networks; we need to isolate servers in the airlock state from other servers in the same state so that a compromised server cannot infect other un-compromised servers. (\textbf{2}) The secure firmware is executed (if stored in system flash) or provisioned onto the server along with a boot-loader, attestation software agent, and any other related software. With these in place, (\textbf{3}) the Attestation Service attests the integrity of the firmware of this server. Once initial attestation completes, (\textbf{4}) the tenant again employs the Isolation Service to move the server from the airlock network. If firmware attestation failed (\textbf{5}) it is moved into the Rejected Pool, isolated from the rest of the cloud; if attestation was successful, the server is made part of the tenant's enclave by connecting it to the tenant networks. In order to make use of the server, further provisioning is required (\textbf{6}) so the tenant again uses the Provisioning Service to install the tenant operating system and any other required applications.

\subsection{Use Cases}

\begin{figure}[t]
	\centering
	\includegraphics[width=0.8\linewidth]{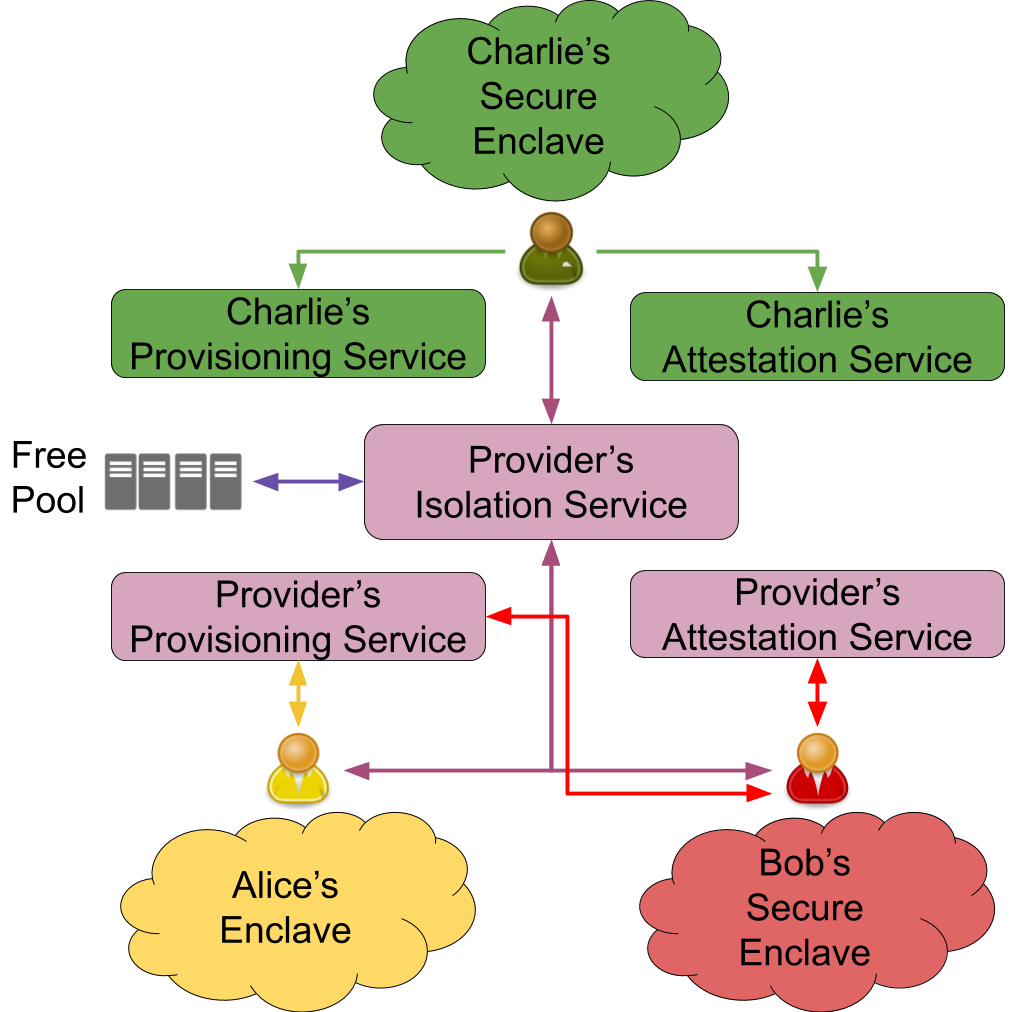} 
\fixfig
	\caption{\bolt\ deployment examples; purple boxes are provider-deployed and greens are tenant-deployed. Alice and Bob trust the provider-deployed infrastructure, while security-sensitive Charlie deploys its own.}
\fixfig
\vspace{-4mm}
	\label{fig:complex}
\end{figure}

Figure~\ref{fig:complex} demonstrates the flexibility of \bolt\ using three examples of users, namely; 1) Alice, a graduate student, who wants to maximize performance and minimize cost and does not care about security, 2) Bob, a professor, who does not trust other tenants (e.g., graduate students) but is willing to trust the provider, and 3) Charlie, a security-sensitive tenant, who not only does not trust other tenants but wants to minimize his trust in the provider.

Alice and Bob are willing to trust the provider's network isolation and storage security, and do not need to employ runtime encryption and will not incur its performance burden; nor will they need to expend the effort to deploy and manage their own services\footnote{Or mismanage, a more significant risk for less security-literate users.}. Alice, further, uses scripts that do not even bother using the provider's attestation service, further improving the speed that she can start up servers as well as her costs if the provider charges her for all the time a server is allocated to her.


Security-sensitive tenant Charlie deploys his own, potentially modified, provisioning and attestation service.  He does not have to rely on the provider's network isolation to protect his confidentiality and integrity but can implement runtime protections such as network and disk encryption.  Moreover, the attestation service can be used not only to protect him from previous tenants, but also to maintain a whitelist of applications and configuration, and to quickly detect any compromises in an ongoing fashion. The one area where \bolt\ requires Charlie to trust the provider is for protecting against denial of service attacks since only the provider can deploy the isolation service that allocates servers and controls provider switches.  Trusting a provider, in this case, is unavoidable with current networking technology, as the provider controls all networking to the datacenter.

In addition to the cloud use cases, \bolt\ was designed to be flexible enough to handle the use case of co-location facilities~\cite{mghpcc_about:2018,NWRDC:2018, equinix_privatecloud:2018} where the datacenter tenants temporarily ``loan'' computers to each other to handle fluctuations in demand; and this use case is, in fact, the primary one for which \bolt\ is going into production currently.  In this case, a single party may be both provider and tenant.
As an example, one party might have a high demand on their HPC cluster, while another party has spare capacity in their IaaS cloud; the isolation service from the second party (the provider) could be used to provision servers for loan to the first party, with attestation and provisioning services (including provisioning-associated storage) provided by the first party (the user).

Since the different \bolt\ services are independent, being orchestrated by tenant scripts, it is straightforward for a tenant to use capacity from multiple isolation services.  The attestation of \bolt\ is important to enable supporting untrusted environments (e.g., research testbeds) alongside production services.  For tenants that use the standard \bolt\ provisioning service, the use of network mounted storage by \bolt\ enables them to use their own storage for persistence, making storage encryption unnecessary.  Because \bolt\ enables tenants to deploy their own provisioning service, some tenants can use custom provisioning services which install to local storage.\footnote{In this case, provisioning time is much larger and tenants are responsible for scrubbing the local disk.}  When using their own infrastructure, the tenant and provider are in the same organization.   In this case, tenants trust the provider, and hence network encryption is unnecessary.  Tenants are willing to make agreements with trusted partners from whom they will be using servers; trusting the partner's isolation service makes network encryption unnecessary for communication with servers obtained from it.

\section{Implementation}\label{impl_sec}

We describe our implementation of the Isolation Service (\hil~\cite{hil}), Firmware (\linuxboot~\cite{linuxboot_2018}), Attestation Service (\keyl~\cite{keylime}), and Provisioning Service (\bmi~\cite{bmi}), and explain how they work together as \bolt. All of these constituent services of \bolt\ are open-source packages and can be modified by tenants or providers to meet their specific requirements.


\textbf{\lhil:} The fundamental operations \lhil\ (\hil) provides are (i) allocation of physical servers, (ii) allocation of networks, and (iii) connecting these servers and networks. A tenant can invoke \hil\ to allocate servers to an enclave, create a management network between the servers, and then connect this network to any provisioning tool (e.g.,~\cite{emulab, Ironic:2018, bmi, canonical_maas2018}). It can also let tenants create networks for isolated communication between servers and/or attach those servers to public networks made available by the provider. \hil\ controls the network switches of the cloud provider and provides VLAN-based~\cite{vlan} network isolation mechanism.  \hil\ also supports a simple API for Baseboard Management Controller (BMC) operations like power cycling servers and console access; ensuring that users cannot attack the BMC. \hil\ \changed{cannot be deployed by tenants and must be } deployed by the provider and is the only component shared by tenants, that is not attested to.  In our effort to minimize this TCB  we have worked hard to keep \hil\ very simple (approximately 3000 LOC).

Because the provider is trusted for physical isolation and security, it also acts as the source of truth for information on servers in two ways.  First, it maps each server's \hil\ identity to a TPM identity by exporting the TPM's public Endorsement Key (EK) through administrator-modifiable metadata per server, ensuring that the tenant is able to confirm that the tenant she received is indeed the one she reserved thus protecting the tenant from any server spoofing attack. Second, \hil\ exposes the provider-generated whitelist of TPM PCR measurements, i.e., ones that relate to the platform components like BIOS/UEFI firmware and firmware settings.

\textbf{\llinuxboot:} \linuxboot~is our firmware implementation and bootloader replacement. It is a minimal reproducible build of Linux that serves as an alternative to UEFI and Legacy BIOS.  \linuxboot\ retains the vendor PEI (Pre-EFI environment) code as well as the signed ACM (authenticated code modules) that Intel provides for establishing the TEE (trusted execution environment).  \linuxboot\ replaces the DXE (Driver Execution Environment) portion of UEFI with open source wrappers, the Linux Kernel, and a flexible initrd based runtime.  Advantages over stock UEFI include: 1) \linuxboot's open-source Linux devices drivers and filesystems have had significantly more scrutiny than the UEFI implementations, 2) its deterministic build enables easy remote attestation with a TPM; a tenant can independently confirm that the firmware on a server corresponds to source code that they compile themselves, 3) it can use any Linux-supported filesystem or device driver, execute Linux shell scripts to perform remote attestation over secure network protocols and mount encrypted drives, simplifying integration into services like \bolt, 4) it is significantly faster to POST than UEFI;  taking 40 seconds on our servers, compared to about 4 minutes with UEFI.

\changed{We chose \linuxboot\ over alternatives like Tianocore~\cite{tianocore:2019} -- an open source implementation of UEFI because unlike Tianocore it does not depend on hardware drivers provided by motherboard vendors. In addition to the driver dependency Tianocore also needs support of Firmware Support Package (FSP) from processor vendors which are closed source binaries or independent softwares like coreboot~\cite{coreboot:2019, coreboot_payload:2019} to function as a complete bootable firmware.
\linuxboot\ does use FSP however Heads which is our flavor of \linuxboot\ is able to establish root of trust prior to executing FSP thus ensuring that FSP blob is measured into TPM PCR's. This protects from attacks that involve replacing a measured FSP with a malicious FSP. Additionally, while \linuxboot\ and Tianocore both are open source projects, \linuxboot\ is based on Linux, a much more mature and widely used system with battle tested code.}

\changed{ We have modified \linuxboot\ such that it scrubs memory before a tenant can use a server; a tenant that attests that \linuxboot\ is installed is guaranteed that subsequent tenants will not gain control until the memory has been scrubbed since the only way for the provider, or another tenant, to gain control (or reflash the firmware) is to power cycle the machine which will ensure that \linuxboot\ is executed.} Scripts integrated with \linuxboot\ download the attestation service's client side agent, download and kexec a tenant's kernel (only if attestation has succeeded), and obtain a key from the attestation service to access the encrypted disk and network.


\textbf{\lkeyl:} \keyl\ is our remote attestation and key management system. It is divided into four major components: Registrar, Cloud Verifier, Agent, and Tenant.  The registrar stores and certifies the public \textit{Attestation Identity Keys (AIKs)} of the TPMs used by a tenant; it is only a trust root and does not store any tenant secrets.  The Cloud Verifier (CV) maintains the whitelist of trusted code and checks server integrity.  The Agent is downloaded and measured by the server (firmware or previously measured software) and then passes quotes (i.e., TPM-signed attestations of the integrity state of the machine) from the server's TPM to the verifier. The Tenant starts the attestation process and asks the Verifier to verify the server.  The Registrar Verifier and Tenant can be hosted by the tenant outside of the cloud or could be hosted on a physical system in the cloud.
\keyl\ delivers the tenant kernel, initrd and scripts to the server (after attestation success) using a secure connection between the \keyl\ CV and \keyl\ agent. The script is executed by the agent to 1) make sure the server is on the tenant's secure network and 2) kexec into tenant's kernel and boot the server.

For tenants that do not trust the provider, \keyl\ supports automatic configuration for Linux Unified Key Setup (LUKS)~\cite{luks:2018} for disk encryption and IPsec for network encryption using keys bootstrapped during attestation and bound to the TPM hardware root-of-trust. Also, \keyl\ integrates with the Linux Integrity Measurement Architecture (IMA)~\cite{ima} to allow tenants to continuously attest that a server was not compromised after boot.  IMA continuously maintains a hash chain rooted in the TPM of all programs, libraries, and critical configuration files that have been executed or read by the system.  The CV checks the IMA hash chain regularly at runtime to detect deviations from the whitelist of acceptable hashes.

\textbf{\lbmi:} The fundamental operations provided by the \lbmi\ (\bmi) are: (i) disk image creation, (ii) image clone and snapshot, (iii) image deletion, and (iv) server boot from a specified image. Similar to virtualized cloud services, \bmi\ serves images from remote-mounted boot drives, with server access via an iSCSI (TGT~\cite{tgt}) service managed by \bmi\, and back-end storage in a Ceph~\cite{ceph} distributed storage system.  When the server network-boots, it only fetches the parts of the image it uses (less than 1\% of the image is typically used), which significantly reduces the provisioning time~\cite{bmi}.   \bmi\ allows tenants to run scripts against a \bmi-managed filesystem which we use to extract boot information (kernel, initramfs image and kernel command lines) from images so that they could be passed to a booting server in a secure way via \keyl.

\textbf{Putting it together:} The booting of a server is controlled by a Python application that follows the sequence of steps shown in Figure~\ref{fig:simple}.  For servers that support it, we burn \linuxboot\ directly into the server's SPI flash.  Figure~\ref{fig:simple} shows another case where we download \linuxboot's runtime (Heads) using iPXE and then continue the same sequence as if \linuxboot\ was burned into the flash.  We have modified the iPXE client code to measure the downloaded \linuxboot\ runtime image into a TPM platform configuration register (PCR) so that all software involved in booting a server can be attested.  When servers pass attestation, the \keyl\ Agent downloads an encrypted zip file containing the tenant's kernel, initrd, and a script from \keyl\ server and unzips them.  The zip file also includes the keys for decrypting the storage and network.  After a server is moved (using \hil) to the tenant's enclave, the \keyl\ Agent runs the script file. The script stores the cryptographic keys into an initrd file to pass it to the tenant's kernel and then kexecs into the downloaded kernel. After it boots, the kernel uses the keys from the initrd file to decrypt the remote disk and encrypt the network.

\changed{\keyl~\cite{keylime} and \linuxboot~\cite{linuxboot_2018} were previously created in part by authors of this paper, and modified as discussed above. While previously published, \hil~\cite{hil} and \bmi~\cite{bmi} were designed with the vision of integrating them in the larger \bolt\ architecture described in this paper.}




\begin{figure*}[h]
\centering
 \begin{subfigure}{0.32\textwidth}
  	\centering
	\includegraphics[width=0.9\textwidth]{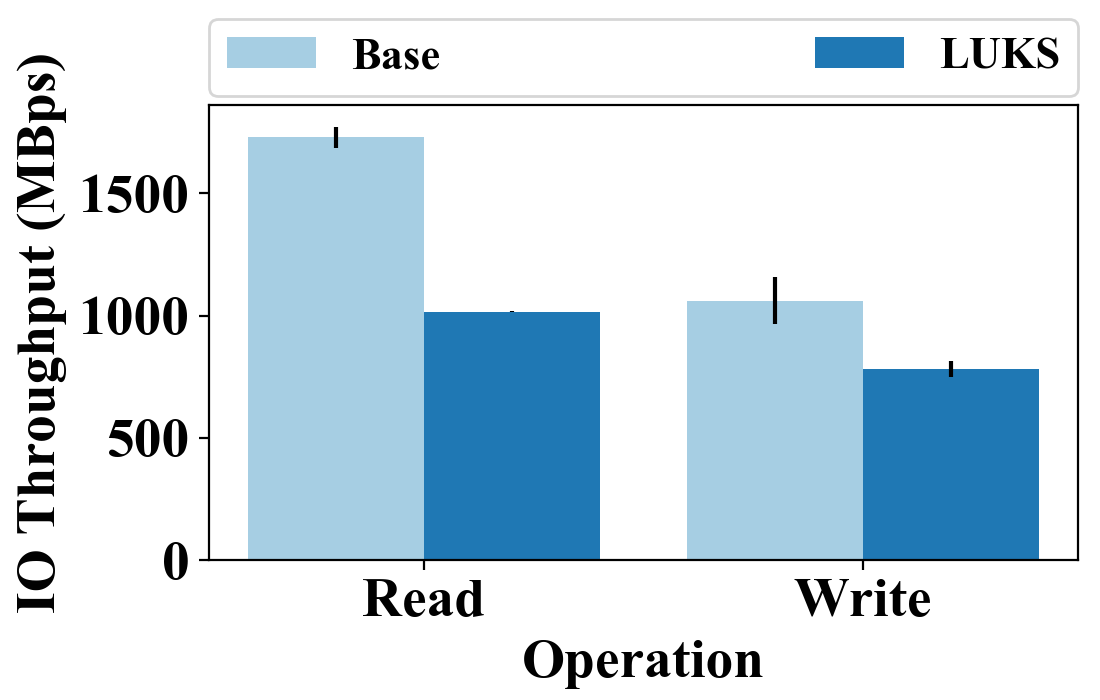}\hfill
	\caption {LUKS overhead on RAM disk}
	\label{fig:luks_ram}
 \end{subfigure}
 \begin{subfigure}{0.32\textwidth}
	\centering
	\includegraphics[width=0.9\textwidth]{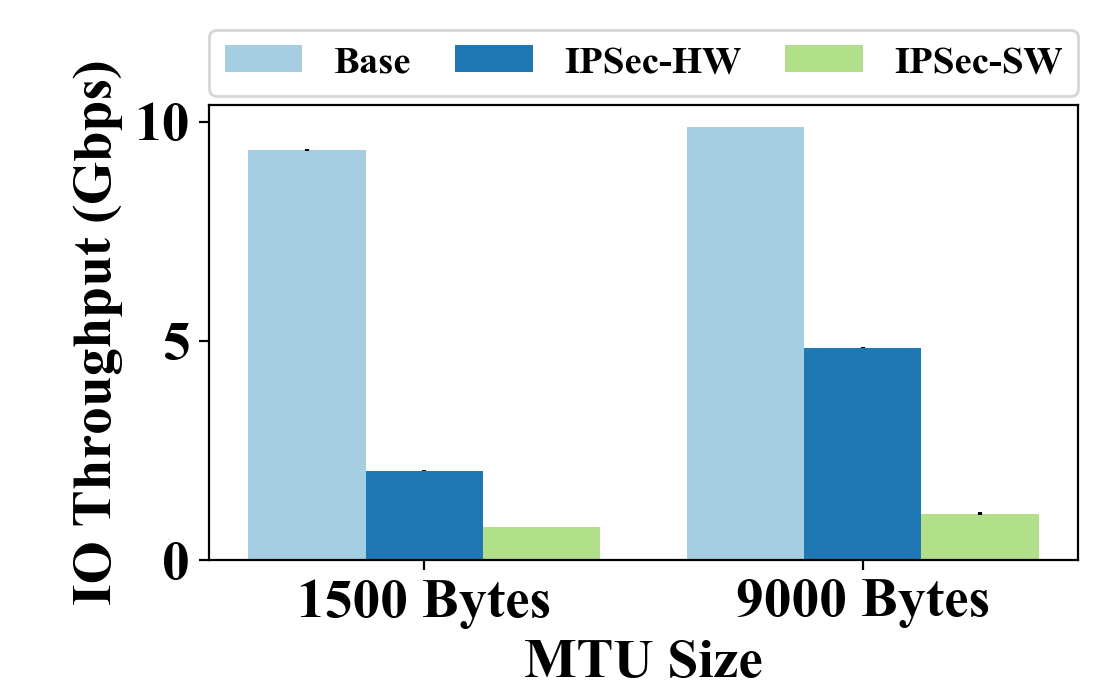}\hfill
	\caption {IPsec overhead}
	\label{fig:ipsec}
 \end{subfigure}
\begin{subfigure}{0.32\textwidth}
	\centering
	\includegraphics[width=0.9\textwidth]{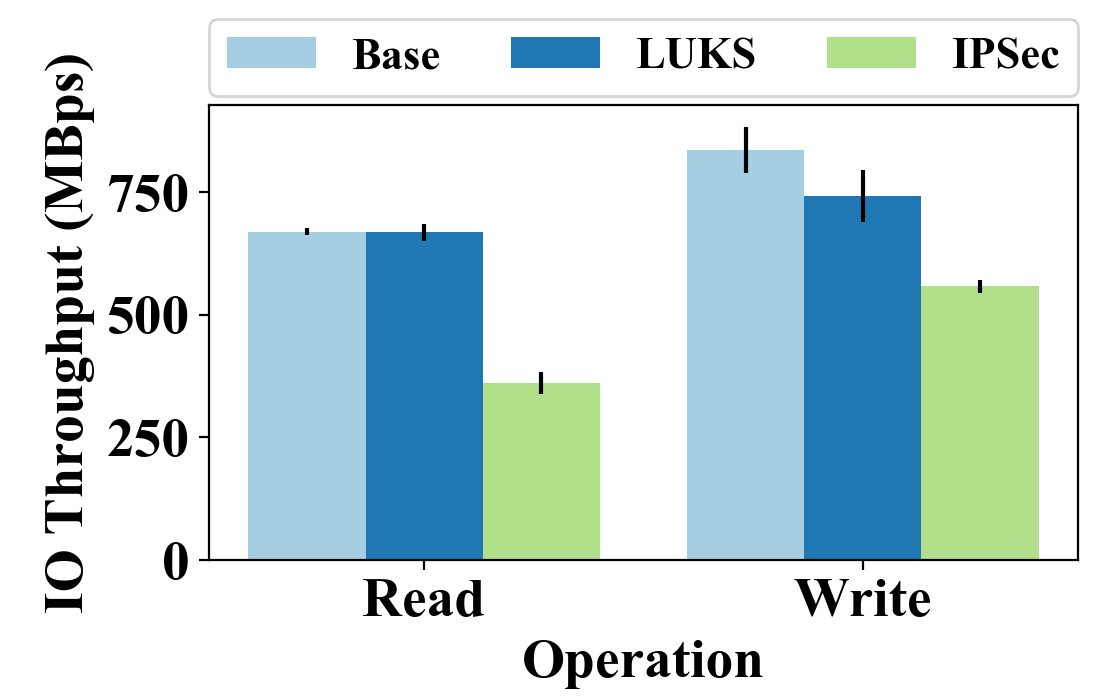}
	\caption {Network mounted storage}
	\label{fig:rbd_overhead}
\end{subfigure}
\fixfig
\caption {Performance Impact of Encryption}
\fixfig
\vspace{-3mm}
\label{fig:components}
\end{figure*}

\section{Addressing the Threat Model}

Here we discuss how, for security-sensitive tenants, \bolt's architecture addresses the threats in the three phases described in Section~\ref{threat}.

\textbf{Prior to occupancy:} We must protect a tenant's server against threats from previous users of the server and isolate it from potential network-based attacks until a server is fully provisioned.  To do this, \bolt\ uses attestation to ensure that the firmware of the server was not modified by previous tenants, and isolates the server in the airlock state (protected from other tenants) until this attestation is complete.  The deterministic nature of \linuxboot\ enables tenants to inspect the source code of the firmware, and ensure that it is trusted, rather than just trusting the provider.  Further, all communication within the networks in the enclave is encrypted, using a key provided by the tenant's attestation service (e.g., \keyl), ensuring that the server will not be susceptible to attacks by other servers as it is provisioned.
Since our current implementation is unable to attest the state of peripheral firmware, there could be malware embedded in those devices that could compromise the node.
Disk and network encryption securely bootstrapped by the TPM mitigate data confidentiality and integrity attacks from malicious peripherals with external access like network interfaces and storage controllers.
System level isolation of device drivers, as in Qubes\footnote{\url{https://www.qubes-os.org/}}, could further be used to mitigate the impact of malicious peripherals mounting attacks against the node~\cite{google0_1}.

\textbf{During occupancy:} We must ensure that the server's network traffic is isolated so that the provider or other concurrent tenants of the cloud cannot launch attacks against it or eavesdrop on its communication with other servers in the enclave.
\hil\ performs basic VLAN-based isolation to provide basic protection from traffic analysis by other tenants.  However, a tenant can choose to both encrypt their network traffic with IPsec and shape their traffic to resist traffic analysis from the provider and not rely on provider's \hil.
\keyl\ securely sends the keys for encrypting networking and disk traffic directly to the node. Disk encryption ensures the confidentiality and integrity of the persistent data even if the storage is under the control of a malicious provider.

Continuous attestation can detect changes to the runtime state of the server (e.g., unauthorized binaries being executed or reboot to an unauthorized kernel) and notify the tenant to take some action to respond.  Response actions include revoking the cryptographic keys used by that server for network/storage encryption, removing it from the enclave VLAN, and immediately rebooting the system into a known good state and scrubbing its memory.
While IMA only supports load/read-time measurement (i.e., hashing) of files on the system as they are used, most existing runtime protection measures like kernel integrity monitoring~\cite{LKIM}, control-flow integrity~\cite{cfi_survey}, or dynamic memory layout randomization~\cite{tasr} are built into either the kernel image/modules, application binaries, or libraries themselves.  \changed{Thus, TPM measurements created by IMA at runtime will demonstrate that those protections were loaded.}

\textbf{After occupancy:} Once a server is removed from a tenant enclave, we must ensure that the confidentiality of a tenant is not compromised by any of its state being visible to subsequent software running on the server.  Stateless provisioning of the servers protects against any persistent state remaining on the server and avoids any reliance on the provider scrubbing the disk if it preempts the tenant.  Further, the tenant can deploy its own provisioning service and ensure that the provider has no access to that storage.  If the tenant requires the use of the local disks for performance reasons (e.g., for big data applications), the server can use local disk encryption with ephemeral keys stored only in memory.  As long as the tenant attests that \linuxboot\ is used, it knows that this firmware will zero the server's memory before another tenant will have the opportunity to execute any code.\footnote{Note that we are assuming here that the provider cannot re-flash the firmware remotely over the BMC.}

\begin{figure*}[t]
  \centering
  \includegraphics[width=0.88\textwidth]{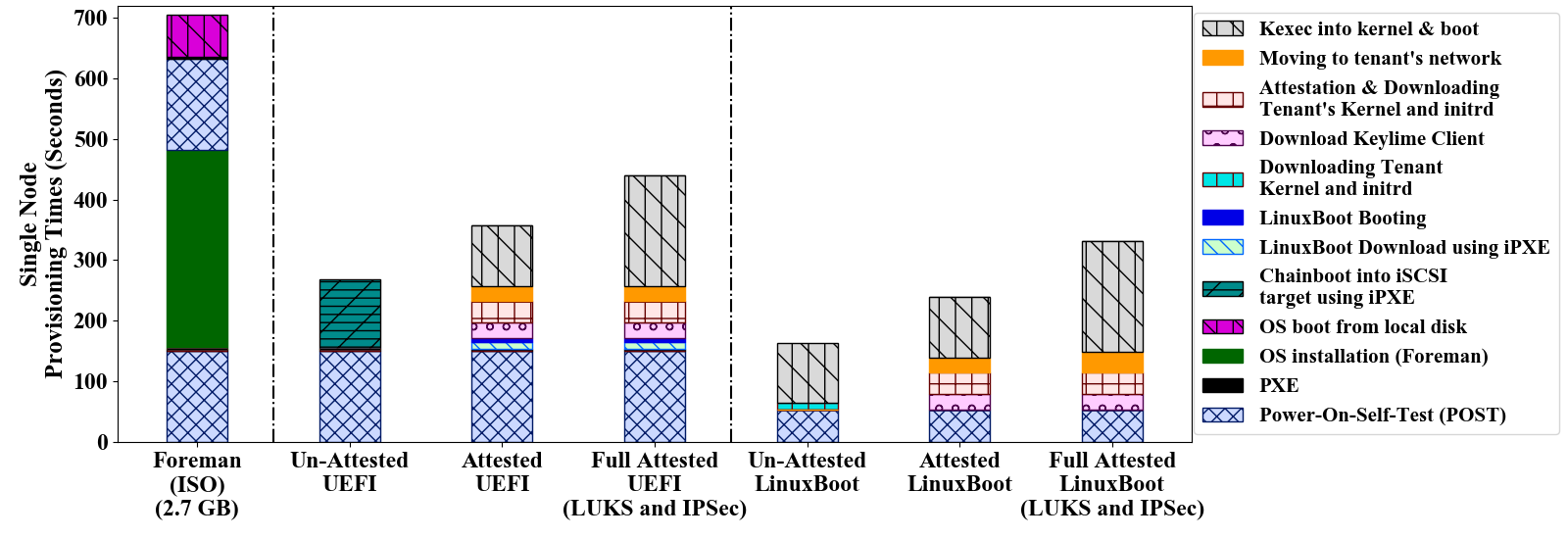}
\fixfig
  \caption {Provisioning time of one server.}
\fixfig
\vspace{-3mm}
  \label{fig:prov}
\end{figure*}

\section{Evaluation}\label{eval_sec}

We first use micro-benchmarks to quantify the cost of encrypted storage and networking on our system, then examine the performance and scalability of the \bolt\ prototype, the cost of continuous attestation and finally the performance of applications deployed using \bolt\ under different assumptions of trust.






\subsection{Infrastructure and methodology}
Single server provisioning experiments were performed on a Dell R630 server with 256 GB RAM and 2 Xeon E5-2660 v3 2.6GHz processors with 10 (20 HT) cores, using UEFI or \linuxboot\ executing from motherboard flash.
All the other experiments were conducted on a cluster of 16 Dell M620 blade servers (64 GB memory, 2 Xeon E5-2650 v2 2.60GHz processors with 8 cores (16 HT) per socket) and a 10Gbit switch. The M620 servers do not have a hardware TPM, so for functionality, we used a software emulation of a TPM~\cite{IBM-tpm1.2:2019}, and for performance evaluation, emulated the latency to access the TPM based on numbers collected from our R630 system.


\hil, \bmi, and \keyl\ servers were run on virtual machines with Xeon E5-2650 2.60GHz CPUs: \keyl\ with 16 vCPUs and 8GB memory; \bmi\ with 2 vCPUs and 8GB, and \hil\ with 8 vCPUs and 8GB RAM. The iSCSI server ran on a virtual machine with 8 vCPUs and 32GB RAM. The Ceph cluster (the storage backend for \bmi\ disk images) has 3 OSD servers (each dual Xeon E5-2603 v4 1.70GHz CPUs, 6 cores each) and a total of 27 disk spindles across the 3 machines. The servers were provisioned with Fedora 28 images (Linux kernel 4.17.9-200) enabled with IMA and version 5.6.3 of Strongswan~\cite{strongswan} for IPsec. IPsec was configured in 'Host to Host' and Tunnel mode. The cryptographic algorithm used was AES-256-GCM SHA2-256 MODP2048. The authentication and encryption were done through a pre-shared key (PSK). IMA used SHA-256 hash algorithm. Cryptsetup utility version 1.7.0 was used to setup disk encryption based on LUKS -- with AES-256-XTS algorithm. Unless otherwise stated, each experiment was executed five times.

\subsection{The cost of encryption}

For security-sensitive tenants that do not trust a provider, they must encrypt the disk and network traffic.   To understand the overhead in our environment, we ran some simple micro-benchmarks.

\textbf{Disk Encryption:} Figure~\ref{fig:luks_ram} shows the overhead of LUKS disk encryption on a Block RAM disk exercised using Linux's ``dd'' command.  While LUKS introduces overhead in this extreme case, we can see that the bandwidth that LUKS can sustain at 1GB for reads is likely to be able to keep up with both local disks and network mounted storage delivered over a 10Gbit network while write performance may introduce a modest degradation at $\sim$0.8GB.

\textbf{Network Encryption:} Figure~\ref{fig:ipsec} shows the overhead of IPsec using Iperf between two servers using both hardware-based Intel AES-NI (IPsec HW) and software-based AES (IPsec SW) and MTU's of 1500 and 9000.  We can see that IPsec has a much larger performance overhead than LUKS disk encryption, with even the best case of HW accelerated encryption and jumbo frames having almost a factor of two degradation over the non-encrypted case \changed{(CPU usage on our infrastructure is between 60\% and 80\% of one processing core for HW accelerated encryption)}.  Additional tuning or specialized IPsec acceleration network interfaces could be used to boost performance~\cite{ipsec_perf}. We use hardware accelerated encryption and jumbo frames for all subsequent experiments.

\textbf{Network mounted storage:} In our implementation we boot servers using iSCSI which in turn accesses data from our Ceph cluster.  In Figure~\ref{fig:rbd_overhead} we show the results of exercising the iSCSI server using ``dd''. \changed {Experimentally, we found that increasing the read ahead buffer size on Linux to 8MB was critical for performance, and we do this on all subsequent experiments (the default size is 128KB).
Since Ceph as the backend storage reads data in 4MB chunks, increasing the read ahead buffer size to 8MB results in higher sequential read performance.}
 As expected we find that LUKS introduces small overhead on writes and no overhead on reads, while IPsec between the client and iSCSI server has a major impact on performance.



\subsection{Elasticity}

Today's bare-metal clouds take many tens of minutes to allocate and provision a server~\cite{Omote:2015}.  Further, scrubbing the disk can take many hours; an operation required for stateful bare metal clouds whenever a server is being transferred between one tenant and another.  In contrast, virtualized clouds are highly elastic; provisioning a new VM can take just a few minutes and deleting a VM is nearly instantaneous.  The huge difference in elasticity between bare-metal clouds and virtualized clouds has a major impact on the use cases for which bare-metal clouds are appropriate.  How close can we approach the elasticity of today's virtualized clouds?  What extra cost does attestation impose on that elasticity?  What is the extra cost if the tenant does not trust the provider and need to encrypt disks and storage?

To understand the elasticity \bolt\ supports, we first examine its performance for provisioning servers under different assumptions of security and then examine the concurrency for provisioning multiple servers in parallel.

\paragraph{Provisioning time:}Figure~\ref{fig:prov} compares the time to provision a server with Foreman (a popular provisioning system)~\cite{foreman:2019} to \bolt\ with both UEFI and \linuxboot\ firmware under 3 scenarios: \emph{no attestation} which would be used by clients that are insensitive to security, \emph{attestation} where the tenant trusts the provider, but uses (provider deployed) attestation to ensure that previous tenants have not compromised the server, and \emph {Full attestation}, where a security-sensitive tenant that does not trust the provider uses LUKS to encrypt the disk and IPsec to encrypt the path between the client and iSCSI server.
There are a number of important high-level results from this figure.  First for tenants that trust the provider, \bolt\ using \linuxboot\ burned in the ROM is able to provision a server in under 3 minutes in the unattested case and under 4 minutes in the attested case; numbers that are very competitive with virtualized clouds.  Second, attestation adds only a modest cost to provisioning a server and is likely a reasonable step for all systems.  Third, even for tenants that do not trust the provider,  (i.e. LUKS \& IPsec) on servers with UEFI, \bolt\ at $\sim$7 minutes is still 1.6x faster than Foreman provisioning; note that Foreman implements no security procedures and is likely faster than existing cloud provisioning systems that use techniques like re-flashing firmware to protect tenants from firmware attacks.

Examining the detailed time breakdowns in Figure~\ref{fig:prov}; while we introduced \linuxboot\ to improve security, we can see that the improved POST time (3x faster than UEFI) on these servers has a major impact on performance. We also see that booting from network mounted storage, introduced to avoid trusting the provider to scrub the disk, also has a huge impact on provisioning time.  The time to install data on to the local disk is much larger for the Foreman case, where all data needs to be copied into the local disk.  In contrast, with network booting, only a tiny fraction of the boot disk is ever accessed.  We also see that with a stateful provisioning system like Foreman,  it needs to reboot the server after installing the tenant's OS and applications on the local disk of the server; incurring POST time twice.  While not explicitly shown here, it is also important to note that with \bolt\ a tenant can shutdown the OS and release a node to another tenant and then later restart the image on any compatible node; a key property of elasticity in virtualized clouds that is not possible with stateful provisioning systems like Foreman.

We show in Figure~\ref{fig:prov} the costs of all the different phases of an attested boot.
With UEFI, after POST, the phases are: (i) PXE downloading iPXE, (ii) iPXE downloading \linuxboot's runtime (Heads), (iii) booting \linuxboot, (iv) downloading the \keyl\ Agent (using HTTP), (v) running the \keyl\ Agent, registering the server and attesting it, and then downloading the tenant's kernel and initrd, (vi) moving the server into the tenant's network and making sure it is connected to the \bmi\ server and finally (vii) \linuxboot\ kexec'ing into the tenant's kernel and booting the server.
In each step, the running software measures the next software and extends the result into a TPM PCR.  Using \linuxboot\ firmware, after POST we immediately jump to step (iv) above.

While the steps for attestation where complex to implement, the overall performance cost is relatively modest, adding only around 25\% to the cost of provisioning a server.\footnote{Moreover, given that performance is sufficient, we have so far made no effort to optimize the implementation.  Obvious opportunities include better download protocols than HTTP, porting the \keyl\ Agent from python to Rust, etc.} This is an important result given a large number of bare-metal systems (e.g. CloudLab, Chameleon, Foreman, \ldots), that take no security measure today to ensure that firmware has not been corrupted.  There is no performance justification today for not using attestation, and our project has demonstrated that it is possible to measure all components needed to boot a server securely. For the full attestation scenarios (UEFI and \linuxboot), two more steps are added to the basic attestation scenarios: (+i) loading the cryptographic key and decrypting the encrypted storage with LUKS (+ii) establishing IPsec tunnel and connecting to the encrypted network. These two steps are incorporated into Kernel boot time in Figure~\ref{fig:prov}. We can see that the major cost is not these extra steps but the slow down in booting into the image that comes from the slower disk that is accessed over IPsec.



\begin{figure}[t]
		\centering
		\includegraphics[width=0.28\textwidth]{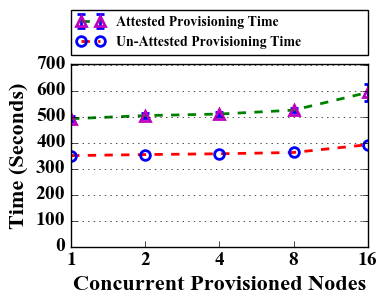}
        \fixfig
	    \caption {\bolt\ Concurrency}
        \fixfig
\vspace{-4mm}
		\label{fig:scaling}
\end{figure}

\paragraph{Concurrency:}Figure~\ref{fig:scaling} shows (with UEFI firmware) how \bolt\ performs, with and without attestation, as we increase the number of concurrently booting servers (log scale).  In both the attested and unattested case performance stays relatively flat until 8 nodes.  In our current environment, this level of concurrency/elasticity has been more than sufficient for the community of researchers using \bolt. There is a substantial degradation in both the attested and unattested case when we go from 8 to 16 servers. In the unattested case, the degradation is due to the small scale Ceph deployment (with only 27 disks) available in our experimental infrastructure. For the attested boot, the performance degradation arises from a limitation in our current implementation where we only support a single airlock at a time; attestation for provisioning is currently serialized.  While this scalability limitation is not a problem for current use cases in our data center, we intend to address it to enable future use cases of highly-elastic security-sensitive tenants; e.g., a national emergency requiring many computers.



\subsection{Continuous Attestation}

\begin{figure}[t]
	\centering
	\includegraphics[width=0.34\textwidth]{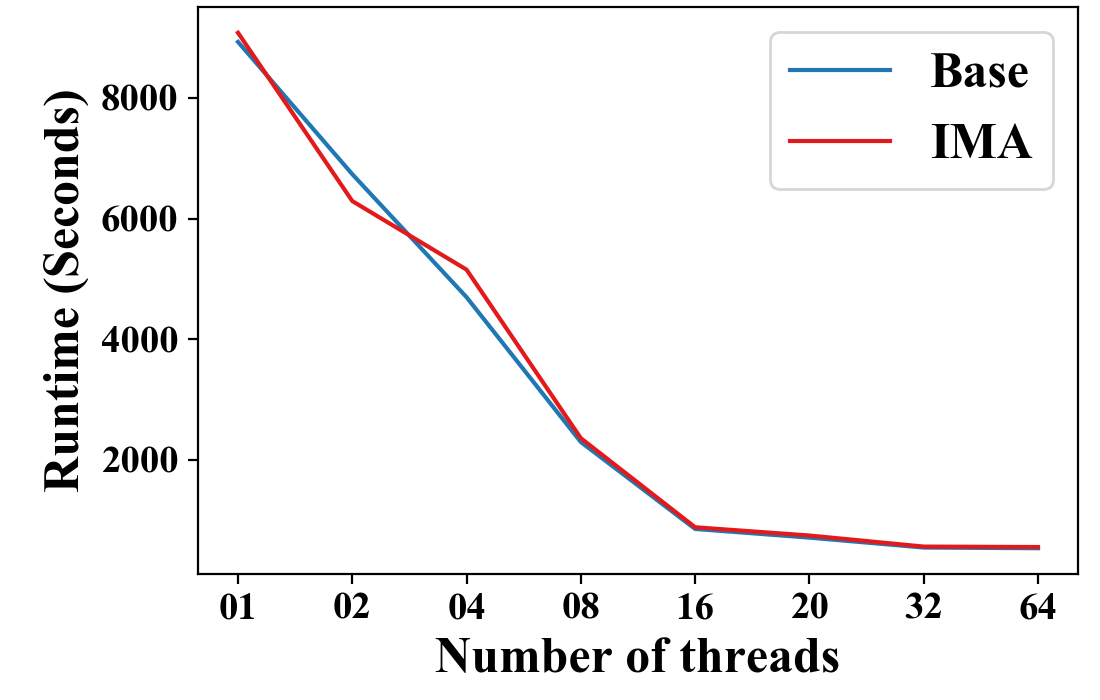}
\fixfig
	\caption {IMA overhead on Linux Kernel Compile}
\fixfig
\vspace{-4mm}
	\label{fig:ima}
 \end{figure}

Once a server has been provisioned, a security sensitive tenant can further use IMA to continuously measure any changes to the configuration and applications.
The \keyl\ Agent will include the IMA measurement list along with periodic continuous attestation quotes. This allows the \keyl\ Cloud Verifier to help ensure the integrity of the server's runtime state by comparing the provided measurement list with a whitelist of approved values provided by the tenant. In the case of a policy violation, \keyl\ can then revoke any keys used for network or disk encryption; essentially isolating the server.
To evaluate IMA performance, we measured Linux kernel 4.16.12 compile time with and without IMA with a different number of processing threads. We use kernel compilation as a test case for IMA because it requires extensive file I/O and execution of many binaries.  The IMA policy we used measured all files that are executed as well as all files read by the \emph{root} user. To stress IMA we ran the kernel compile as root such that all of its activity would be measured.\footnote{This policy and workload are very unlikely to be either useful or manageable from a security perspective.  We used them only as a stress test.} Figure~\ref{fig:ima} shows the results in log scale; even in this unrealistic stress test IMA does not impose a noticeable overhead.

\keyl\ can detect policy violations from checking the IMA measurements and TPM quotes in under one second. To simulate a policy violation, we ran a script on the server without having a record of it in the whitelist, resulting in an IMA measurement different than expected.  This results in \keyl\ issuing a revocation notification for the key of the affected server used for IPsec to the other servers in the system; the entire process takes approximately 3 seconds for a compromised server to have its IPsec connections to other servers reset and be cryptographically banned from the network.


\subsection{Macro-Benchmarks}

Security-sensitive tenants using \bolt\ rely on network and disk encryption to minimize their trust in the provider.  Surprisingly there is little information in the literature what the cost of such encryption is for real applications.  Is the performance good enough that we can tolerate a one-size-fits-all solution and avoid ever trusting the provider? Is the performance so poor that it will never make sense for security-sensitive customers to use \bolt?

\begin{figure}[t]
	\centering
	\includegraphics[width=0.41\textwidth]{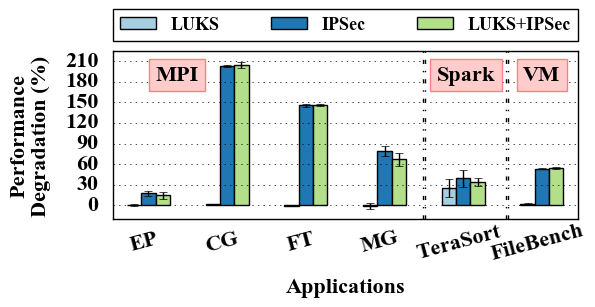}
\fixfig
	\caption {Macro-benchmarks' performance} 
\fixfig
\vspace{-3mm}
	\label{fig:macros}
\end{figure}

Figure~\ref{fig:macros}~(MPI) shows performance degradation results for a variety of applications from the NAS Parallel Benchmark~\cite{bailey1991parallel} version 3.3.1: Embarrassingly Parallel (EP), Conjugate Gradient (CG), Fourier Transform (FT) and Multi Grid (MG) applications class D running in a 16 server enclave. We see overall that these applications only suffer significant overhead for IPsec, ranging from $\sim$18\% for EB, which has modest communication, to $\sim$200\% for CG which is very communication intensive.  These results suggest that there are definitely workloads for which not trusting the provider incurs little overhead.  At the same time, a one-size-fits-all solution is inappropriate; only tenants that are willing to trust the provider, and avoid the cost of encryption, are likely to run highly communication intensive applications in the cloud.

To understand the performance overhead for more cloud relevant workloads, Figure~\ref{fig:macros}~(Spark) shows the performance of Spark~\cite{Spark_2010} framework version 2.3.1 (working on Hadoop version 2.7.7) running TeraSort on a 260GB data set. The experiment is run in parallel in an enclave of 16 servers. TeraSort is a complex application which reads data from remote storage, shuffles temporary data between servers and writes final results to remote storage.  We can see a significant overall degradation, of $\sim$30\% for LUKS+IPsec. While this degradation is significant, we expect that security sensitive tenants would be willing to incur this level of overhead. On the other hand, this overhead is large enough that tenants willing to trust the provider would prefer not to incur it, suggesting that the flexibility of \bolt\ to provide this choice to the tenant is important.

Our last experiment (Figure~\ref{fig:macros}~(VM)) is based on virtualization. An important application of bare metal servers is to run virtualized software (e.g., an IaaS cloud). In this experiment, we installed KVM QEMU version 2.11.2 on a M620 server as the hypervisor. The virtual machine we run on the hypervisor is CentOS 7 with Linux kernel 3.10.0. It has 8 vCPU cores and 32 GB RAM. This is based on the observation ~\cite{vm_config} that 90\% of virtual machines having $\leq 8$ vCPU cores and $\leq 32$ GB RAM. We run Filebench version 1.4.9.1 benchmark~\cite{tarasov2016filebench} on 1000 files with 12MB average size on the virtual machine. We can see that the performance of this benchmark is $\sim$50\% worse in the case of IPsec; a significant performance penalty. While we would expect less of a degradation for regular VMs (rather than ones running a file system benchmark), we can see that a tenant deploying generic services, like virtualization, should be very careful about the kind of workload they expect to use the service.


\section{Related Work}\label{doc:relatedWork}

Our work on creating a secure bare-metal cloud was motivated by a huge body of research demonstrating vulnerabilities due to co-location in virtualized clouds including both hypervisor attacks~\cite{attackcache_2017,coreside_2017,king2006subvirt,Perez-Botero:2013,sze_hardening:2016} and side-channel and cover-channel attacks like the Meltdown and Spectre exploits~\cite{ristenpart2009hey,Kocher2018spectre,Lipp2018meltdown,liu,razavi2016flip}.

There is a large body of products and research projects for bare-metal clouds~\cite{Softlayer:2015,RackSpace:2015,Internap:2015,Packet.net,AWSBare} and cluster deployment systems~\cite{Ironic:2018,canonical_maas2018,ricci_precursors:_2016,xCAT2:2019} that have many of the capabilities of isolation and provisioning that \bolt\ includes.  The fundamental difference with \bolt, as we have explained in ~\cite{bolted-hotcloud}, is that we strongly separate isolation from provisioning and different entities (e.g.\ security sensitive tenants) can control/deploy and even re-implement the provisioning service.  This structuring clearly defines the TCB that needs to be deployed by the provider.

While it is often unclear exactly which technique each cloud uses to protect against firmware attacks, a wide variety of techniques have been used including specialized hardware\cite{titan_gcp:2019, microsoft_cerberus}, using a specialized hypervisor to prevent access to firmware~\cite{fukai_bmcarmor:2017}, and attestation to the provider~\cite{ibmcloud_hardware_2018,inc_oracle_2018}.  In all cases, there is no way for a tenant to programmatically verify that the firmware is up to date and not compromised by previous tenants.  \bolt\ is unique in enabling tenant deployed attestation for bare-metal servers, where the measurement of the firmware and software are provided directly to the tenant.

The static root of trust (SRTM) approach used by \bolt\ requires all software to be measured in an unbroken chain of trust.  It would have been simpler for us to use dynamic root of trust (DRTM), however, DRTM has additional chip dependencies and, more importantly, been shown to be vulnerable to attacks~\cite{wojtczuk2009attacking} and work of Kovah et. al has shown that it can be used as an attack vector itself~\cite{kovah_senter_2014}.

\section{Concluding Remarks}
\label{sec:conc}
We presented \bolt, an architecture for a bare metal cloud that is appropriate for even the most security sensitive tenants; allowing these customers to take control over their own security. The only trust these tenants need to place in the provider is for the availability of the resources and that the physical hardware has not been compromised.  At the same time, by delegating security for security sensitive tenants to the tenants, \bolt\ frees the provider from the complexity of having to directly support these demanding customers and avoids impact to customers that are less security sensitive.

To enable a wide community to inspect the TCB, all components of \bolt\ are open source. We designed \hil, for example, to be a simple micro-service rather than a general purpose tool like IRONIC~\cite{Ironic:2018} or Emulab~\cite{emulab}.  \hil\ is being incorporated into a variety of different use cases by adding tools and services on and around it rather than turning it into a general purpose tool. Another key example of a small open source component is \linuxboot. \linuxboot\ is much simpler than UEFI.  Since it is based on Linux, it has a code base that is under constant examination by a huge community of developers.  \linuxboot\ is reproducibly built, so a tenant can examine the software to ensure that it meets their security requirements and then ensure that the firmware deployed on machines is the version that they require.

\bolt\ protects against compromise of firmware executable by the system CPU; however modern systems may have other processors with persistent firmware inaccessible to the main CPU; compromise of this firmware is not addressed by this approach. These include: Base Management Controllers (BMCs)~\cite{moore_2017}, the Intel Management Engine~\cite{newman_2017,ermolov_goryachy_2017,kroizer_2015}, PCIe devices with persistent flash-based firmware, like some GPUs and NICs, and storage devices~\cite{hd_virus}. Additional work (e.g. IOMMU based techniques, disabling the Management Engine~\cite{me_cleaner} and the use of specialized systems with minimum firmware) will be needed to meet these threats.

The evaluation of our prototype has demonstrated that we can rapidly provision secure servers with competitive performance to today's virtualized clouds; removing one of the major barriers to bare metal clouds.
\changed{
We demonstrate that the cost of not trusting the provider (network/storage encryption) and of additional runtime security (continuous attestation) varies enormously depending on the application.  (In fact, we are not aware of other work that has quantified the cost of network encryption, disk encryption, and continuous attestation with modern servers and implementation.)  Results for HPC applications vary from negligible overhead to three times overhead for communication-intensive applications. Clearly the public cloud becomes economically unattractive for applications with three times overhead unless there are no other alternatives. However, we expect that the $\sim$30\% degradation we see for TeraSort is likely representative of many applications today.  Such overheads suggest that the cost of security is modest enough that security-sensitive customers will find value in using cloud resources.  At the same time, the overhead is significant enough that the flexibility of \bolt\ that enables tenants to just pay for the security they need is justified.}
One surprising result is that our secure firmware, \linuxboot\ achieves dramatically better POST time than existing firmware; this is one of the few times in our experience that additional security comes with performance advantages.

\section{Acknowledgment}
\changed{We would like to acknowledge the feedback of the anonymous reviewers and our shepherd, Dr. Nadav Amit.
We would like to thank Red Hat, Two Sigma and NetApp, the core industry partners of Mass Open Cloud (MOC) for supporting this work.
This project involved extensive efforts over many years to integrate all the components together. We gratefully acknowledge Jason Hennessey, Gerardo Ravago, Ali Raza, Naved Ansari, Kyle Hogan, and Radoslav Nikiforov Milanov for their significant contributions in development and their assistance in the evaluations.
Partial support for this work was provided by the USAF Cloud Analysis Model Prototype project, National Science Foundation awards CNS-1414119, ACI-1440788 and OAC-1740218.}


{

\bibliography{refs,not-anon,db,bolted}
\bibliographystyle{abbrv}
}

\end{document}